\documentclass[a4paper,11pt]{article}
\pdfoutput=1

\usepackage[utf8]{inputenc}
\usepackage{uniinput}
\usepackage{jheppubnew}
\usepackage{bm}
\usepackage{subfig}
\usepackage{caption}
\usepackage{float}

\newcommand{\be}{\begin{equation}}
\newcommand{\ee}{\end{equation}}

\title{An effective three-dimensional de Sitter cosmology in string theory}
\author{Filip Landgren}
\affiliation{Department of Physics and Astronomy, Uppsala University, Box 516, SE-75120, Uppsala, Sweden}

\emailAdd{filip.landgren.8515@student.uu.se}

\begin{document}

\abstract{Motivated by the lack of consensus on whether or not de Sitter (dS) lies in the Swampland, we use a recently developed braneworld construction and known Anti-de Sitter (AdS) vacua to compute an explicit effective dS cosmology in three dimensions. We consider a non-perturbative AdS$_4$ vacuum decaying to another lower AdS$_4$ vacuum via bubble-nucleation. We also consider the more speculative case where a dS$_4$ decay to a Minkowski$_4$. The Israel junction conditions are solved across the bubble and we obtain the Friedmann equations from which the cosmological constants can be read off,  in the respective cases. The cosmological constants are computed in a flux background yielding small positive values admitting a dS cosmology. However, we find that an energetically viable model in the AdS to AdS case requires more fine-tuning.
}

\maketitle

\section{Introduction}
\begin{figure}[H]
\centering
\includegraphics[angle=0, width=0.38\linewidth]{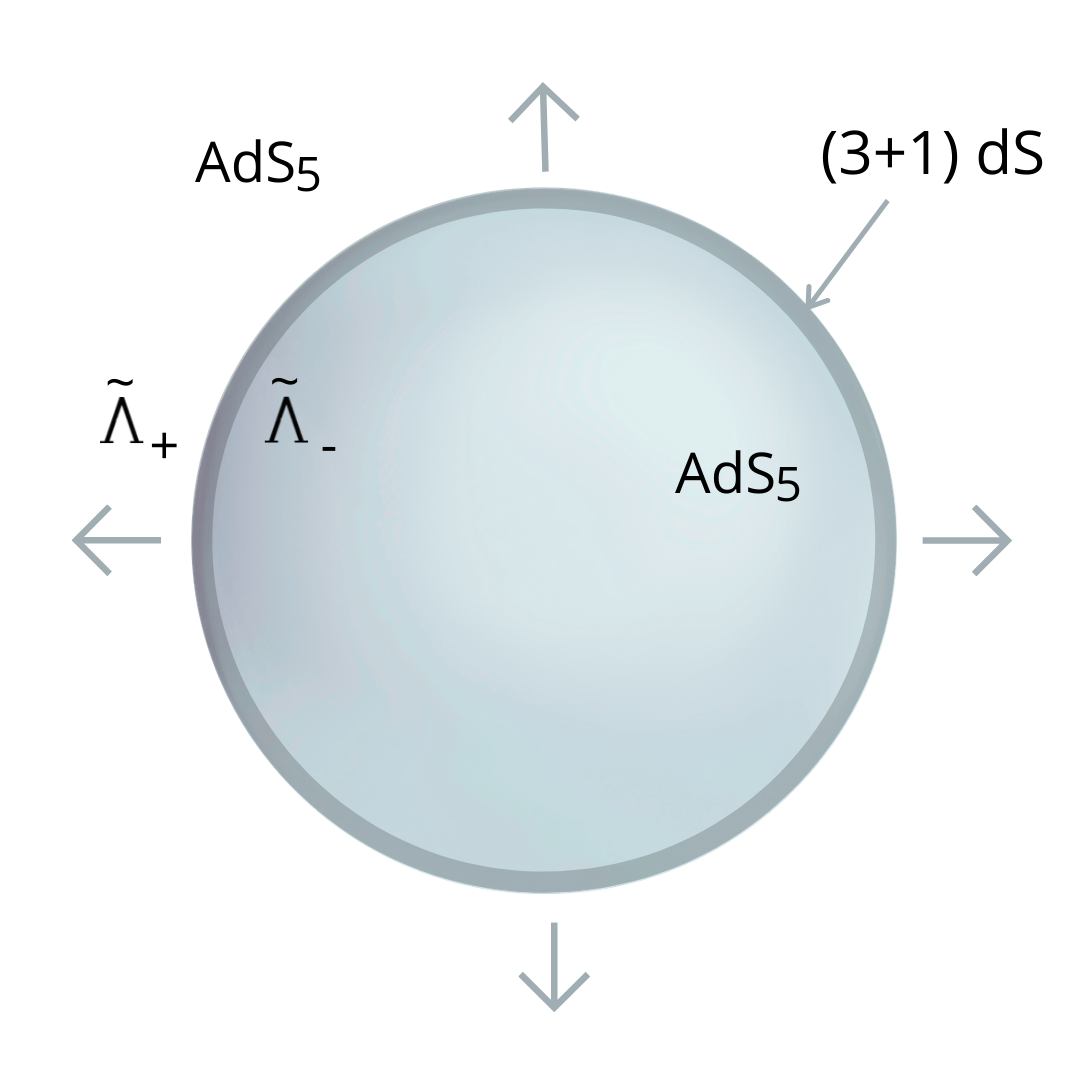}
\caption{An AdS$_5$ vacuum
decaying to another AdS$_5$ vacuum with with more negative
vacuum energy via bubble-nucleation of a true vacuum.}
\label{bubble}
\end{figure}

Compactifying a higher dimensional string theory to a lower dimensional theory typically results in a negative minimum for the scalar potential. The general approach to find de Sitter (dS) has been to find a proper way of "uplifting" an Anti de Sitter (AdS) potential to yield a dS vacuum. Despite the promising results from KKLT-type scenarios \cite{Kachru_2003} and the Large Volume Scenario (LVS)  \cite{Anguelova_2009}, finding a dS vacuum remains an open problem. In part because the AdS vacua used as starting point are often non-SUSY and thus believed to be unstable \cite{ooguri2017nonsupersymmetric}. This means that they will decay perturbatively or non-pertubatively. The continuing struggle of finding a dS vacuum led to the dS Swampland land conjecture stating that no such vacuum should exist in a UV self consistent string theory. In short, the Swampland refers to the consistent quantum effective field
theories that cannot be UV embedded in a theory of quantum gravity \cite{Palti_2019}.  However, in recent work by Banerjee, Danielsson, Dibitetto, Giri and Schillo \cite{Banerjee_2018}, a new and rather different way out of the Swampland was proposed. They state that these vacuum-instabilities are no longer problematic, but rather a necessary ingredient to find a dS cosmology. The idea is that our universe is living on the boundary of a (3+1)-d bubble expanding in a (4+1)-d AdS spacetime. Furthermore, a non-SUSY AdS$_5$ spacetime decay to another,  possibly SUSY, AdS$_5$ spacetime with a lower vacuum energy via the nucleation of a bubble of true vacuum, as illustrated in fig.\ref{bubble}. 

The Israel junction conditions can then be solved across the boundary to compute the Friedmann equations, cosmological constant (cc) and other cosmological properties on the bubble. 

It is important to distinguish this braneworld model with that of Randall-Sundrum (RS) which glues together two insides of AdS spacetimes across a brane so that the spacetime on the brane becomes Minkowski \cite{Randall_1999}, \cite{Randall_19991}.  The model used in this paper deals with both an inside and an outside of a bubble which automatically leads to expansion and dS on the bubble that forms via an instanton with a finite Euclidean action.  Furthermore, gravitational modes are not localized in this picture but generated effectively due to backreaction on the background. This is in contrast to the RS braneworld model where a 4-d gravity is localized on the bubble.

In addition to the AdS to AdS case, we will for completion and curiosity consider the case where we have a 4-d dS spacetime decaying to a 4-d Minkowski spacetime. This scenario is much more speculative as a 4d dS vacuum is required, which per our discussions is not a trusted construction.

The spacetime the bubble is propagating in is refereed to as the bulk spacetime. The reason we work in a 4-d bulk spacetime is because known vacua from KPV, KKLT and LVS can be used to explicitly compute a 3-d effective cosmology which is an equally important example as a higher dimensional cosmology, as far as the Swampland discussion is concerned.
\newpage
\section{The KPV construction}
In this section we briefly review relevant parts of the results of Kachru, Pearson and Verlinde (KPV) that offered a vital ingredient to constructing a dS vacuum \cite{Kachru:2002gs}. They found that placing $p$ anti-branes at the tip of the warped conifold in the KS geometry will contribute with a positive term to the potential energy which enable the 'uplifting-mechanism' of an AdS vacuum as explained below.  The warped geometry of the conifold is described by
\begin{equation}
    \epsilon^2 = \sum_i^4 z_i^2
\end{equation}
where $z_i$ labels the coordinates on $\mathbb{C}_4$.
Consider placing $p<<M K$ number of $\overline{D3}$-branes on the tip of the deformed throat. We will have $M$ units of $F_3$ flux on the three-sphere $S^3$, or A-cycle, as well as $K$ units of $H_3$ flux on the dual B-cycle satisfying the quantization conditions \cite{Kachru:2002gs}:
\begin{equation}\label{12}
\int_A F_3 = 4 \pi^2 \alpha' M, \quad \int_B H_3 = - 4\pi^2 \alpha' K.
\end{equation}
A change in $F_3$ or $H_3$ must be accompanied by a change in the net number of brane charges to satisfy the tadpole condition given by
\begin{equation}\label{tadpole}
      \frac{\chi(X)}{24} = Q_3 + \frac{1}{2 (8 \pi G_{10})^2 T_3} \int_M H_3 \wedge F_3.
\end{equation}
Here $T_3$ is the tension of the $D3$-brane and $\chi(X)$ is the Euler characteristics of the manifold $X$. The right-hand side of this equation counts the net $D3$ charge from transverse fluxes and branes in the CY manifold. The left-hand side accounts for D3-charges of 7-branes in F-theory compactifications and is given by the Euler number of the F-theory fourfold \cite{Giddings_2002}. The tadpole condition plays a crucial role below when adding $\overline{D3}$-brane and fluxes to our system.

Now consider the case where the $\overline{D3}$-branes polarize into NS5-branes. The effective action for the NS 5-branes in type IIB theory is only understood from S-duality of the $D5$-brane action. S-duality maps the type IIB theory to itself and exchanges weak and strong coupling. The D5-brane action is however only valid in a weak coupling regime which should also be true for the NS 5-brane action. But via S-duality, the coupling becomes strong  and outside its regime of validity. More recent papers consider the case where $\overline{D3}$ branes get polarized into D5 branes instead to avoid this issue (see e.g. \cite{gautason2016inflation}).  The action for the $D5$ brane have a different tension and couple to different fluxes in the Chern-Simons term but look very similar to that of the NS 5-branes. Here we will stick with NS 5-branes since the results and intuition obtained from the analysis has been proven to survive this subtlety \cite{Bertolini_2015}.


Due to the presence of $F_5$ fluxes the anti-branes will migrate to the tip of the throat where the metric is given by \cite{dorronsoro2017explicit}
\begin{equation} \label{13}
    ds^2 = \frac{\epsilon^{4/3}}{\alpha' g_s M} dx_\mu dx^\mu + \alpha' g_sM b_0^2( \frac{1}{2}dr^2 + d\Omega_3^2 + r^2 d\Tilde{\Omega}_2^2)
\end{equation}
where $b_0^2 \approx 0.93266$, $\mu$ runs from [0,3] and $r \rightarrow 0$.  The NS 5-brane wraps a $S^2$ inside a finite $S^3$ at an angle $\psi$ which also specifies the radius of the $S^3$: $d\Omega_3 = d\psi^2 + \text{sin}^2(\psi) d\Omega_2$.  The metric of the NS-5 brane can thus be written as
\begin{equation}\label{14}
    ds^2_{NS5} = b_0^2 g_s M \alpha' [dx_\mu dx^\mu + d\psi^2 + \text{sin}^2(\psi) d\Omega_2^2].
\end{equation}
The probe action of the NS-5 brane due to the polarization of $p$ $\overline{D3}$-branes reads \cite{Giddings_2002}:
\begin{equation}\label{15}
    S = \frac{-\mu_5}{g_s} \int d^6 \xi [-\text{det}(G_\parallel)\text{det}(G_\perp + 2 \pi g_s\mathcal{F})]^{1/2} - \mu_5 \int B_6.
\end{equation}
Here $\mathcal{F} = 2 \pi \sqrt{\alpha'}F_2 - C_2$ and $F_2=dA$ is the world-volume field strength.  $G_\parallel$ is the metric along $\psi$ and the non-compact direction and $G_\perp$ is the induced metric along the $S^2$. $B_6$ is defined via $H_7 = dB_6 + \ldots$.
In terms of the polar angle, the action can be expressed as:
\begin{align}\label{22}
S&= \int d^4 x \sqrt{-\text{det}G_\parallel} \mathcal{L}(\psi), \\
\mathcal{L}(\psi) &= \frac{\mu_3 M}{g_s} \Big(V_2(\psi)\sqrt{1-\dot{\psi}^2} - \frac{1}{2\pi}(2\psi -\text{sin}(2\psi)) \Big).
\end{align}
Introducing the canonical momentum, $P_\psi$, conjugate to $\psi$ the Hamiltonian density, $\mathcal{H}$, can be written as \cite{Kachru:2002gs}:
\begin{equation}\label{24}
\mathcal{H}(\psi, P_\psi) = -\frac{\mu_3 M}{2 \pi g_s}(2 \psi - \text{sin}(2\psi) + \sqrt{\Big(\frac{\mu_3 M}{g_s} V_2(\psi) \Big)^2 + P_\psi}.
\end{equation}
Now, the effective potential is given by
\begin{equation}\label{25}
    V^{KPV} = \mathcal{H}(\psi, P_\psi = 0) = \frac{\mu_3}{M} \Big(V_2(\psi) - \frac{1}{2\pi}(2 \psi -\text{sin}(2\psi) \Big)
\end{equation}
which for small values of $\psi$ can be expanded to 
\begin{equation}\label{26}
    V^{KPV} \approx \frac{\mu_3 M}{ \pi g_s} \Big(\frac{p}{M} -\frac{4}{3 \pi} \psi^3 + \frac{b_0^4M}{2 \pi^2 p} \psi^4  \Big)
\end{equation}
with a minimum at $\psi_{min} = \frac{2\pi p}{b_0^4 M}$. The potential at this point reads 
\begin{equation}\label{27}
    V_{min}^{KPV} \approx \frac{\mu_3 p}{g_s} \Big(1-\frac{8\pi^2 p^2}{3b_0^{12}M^2} \Big).
\end{equation}
As KPV found, this analysis is only valid for $\frac{p}{M} \leq 8 \% $ since the slope of the effective potential will otherwise be negative for all values of $\psi$. Furthermore, the anti-branes
are perturbatively unstable for $\frac{p}{M} \geq 8 \% $ and the system classically approaches a
SUSY-vacuum.

A key observation from the effective potential (\ref{25}) is that the
difference in vacuum energy between the south and north-pole is equal to twice the tension
of the a $\overline{D3}$ brane:
\begin{equation}\label{29}
    V^{KPV}(0) - V^{KPV}(\pi) = 2\frac{p\mu_3}{g_s}.
\end{equation}
We can think about this as a comparison between the non-SUSY model with the case where  $\overline{D3}$-branes are replaced by $D3$-branes which preserve SUSY and must thus have a vanishing vacuum energy. To get the true potential, one must therefore add back $p$ $D3/\overline{D3}$ pairs that have vanishing charge and total tension $\frac{2p\mu_3}{g_s}$.  From the tadpole condition (\ref{tadpole}), the true potential then becomes
\begin{equation}\label{30}
    V_{tot}^{KPV} = V^{KPV} + \frac{p \mu_3}{g_s}.
\end{equation}
Hence, the $\overline{D3}$-branes provides a positive contribution to the overall potential.

In addition to the field $\psi$, the string compactification generally generates lighter moduli fields. Giddings, Kachru, and Polchinski (GKP) stabilized all moduli fields except the Kähler modulus \cite{Giddings_2002}, which we denote $u(x)$, that determines the volume of the internal 6-d space. By neglecting backreaction from branes and fluxes the 10-d string frame metric takes the form
\begin{equation}\label{31}
    ds_{10}^2 = g_{\mu \nu} dx^\mu dx^\nu + e^{2u(x)}g_{i, \bar{j}} dz^i d\bar{z}^{\bar{j}}
\end{equation}
where $g_{i, \bar{j}}$ is the Ricci-flat metric. The 4-d effective action thus becomes \cite{Kachru:2002gs}
\begin{equation}\label{32}
    S= \frac{1}{(8 \pi G_4)^2} \int d^4 x\sqrt{-\Tilde{g}_4} \Big( \Tilde{R}_4 - 6 (\partial_\mu u(x))^2 - \frac{\epsilon^{8/3}}{g_sM} e^{-6 u(x)} V_2(\psi) (\partial_\mu \psi)^2 + h_0^4 e^{-12 u(x)} (V_{KPV} + \frac{p \mu_3}{g_s} ) \Big)
\end{equation}
where $h_0$ is the warp factor where the $\overline{D3}$-branes are situated as defined in (\ref{warp}). $\Tilde{g}_4$ and $\Tilde{R}_4$ is the Einstein frame metric and Ricci scalar, respectively. 

The effective 4-d Planck mass is determined by the geometry of the extra dimensions but since there is no string compactification in this setting, we use that the ad hoc relation of the string scale and Planck scale 
\begin{equation}\label{103}
   \alpha' = m_s^{-2} \approx (m_{pl}^{(4)})^{-2} \frac{N}{g_s}
\end{equation}
where $N$ is of order $\sim 10^3$.

 Consider a non-SUSY vacuum with  $p$ $\overline{D3}$-branes. The KPV instanton describes the decay of this vacuum into another lower vacuum via the nucleation of a bubble. A key ingredient in this picture is that the bubble wall is described by a spherical NS 5 domain wall.  When computing the cc in section \ref{cosmologyonexpandingbubble} we can thus use that the bubble tension is given by the product of the NS5-brane tension $\frac{\mu_5}{g_s^2}$ and the volume of the three-sphere $S^3$ with radius $R= b_0 \sqrt{g_s M \alpha' }$ that wraps it:
 \begin{equation}\label{28}
   \mathcal{T}_{NS5} = \frac{\mu_5 2 \pi^2b_0^3(g_s M \alpha ')^{3/2}}{g_s^2}.
\end{equation}

\newpage
\section{Uplifting 4d vacua}

\subsection{The KKLT approach}

KKLT offered one of the most successful prototypes of a dS vacuum in string theory. They used a combination of anti-branes, non-perturbative effects and fluxes to construct a dS vacuum. After freezing all moduli while preserving SUSY, they obtain an AdS vacuum. SUSY is then broken in a controlled manner by adding $\overline{D3}$-branes which uplifts the potential. Adding $\overline{D3}$-branes does not introduce additional moduli as its worldvolume scalars  are frozen by the potential generated by the background fluxes \cite{Kachru:2002gs}.

There is much debate about the validity of uplifting AdS vacua to dS and concerns on weather or not the SUSY breaking is controlled (see e.g \cite{Gao:2020xqh}, \cite{Moritz_2018}, \cite{Frey_2003}, \cite{Landete_2016}).  We will however be interested in the case where the uplifting yields a non-SUSY AdS which will decay and give rise to our braneworld construction. The literature on KKLT is rich and vast so here we will briefly present the relevant background leading up to our construction. 

The total superpotential is given by
\begin{equation}\label{34}
    W= W_0 + \delta W
\end{equation}
where $W_0$ is the tree-level superpotential and $\delta W$ is a non-perturbative correction term required to stabilize the volume modulus. KKLT consider gluino-condensation on the worldvolume of non-Abelian D7-branes as well as wrapped Euclidean D3-branes \cite{Kachru_2003}, giving quantum corrections on the form:
\begin{equation}\label{33}
\delta W=  Ae^{ia \rho}.
\end{equation}
where $A$ and $a$ depend on the energy scales and details of the source of the correction, but is of order $1$ and $\frac{1}{10}$, respectively \cite{Kachru_2003}. 
The volume modulus can be written as $\rho = \tau + i \sigma$. However, KKLT simplifies things by letting the axion, $\tau,$ vanish, which otherwise would contribute with a degenerate prefactor $e^{2ia\tau}$.  Since all moduli has been stabilized except for the volume modulus, the only contributing term from the Kähler potential is
\begin{equation}\label{kähler}
    K = -3ln[-i(\rho - \bar{\rho})] = -3ln[2 \sigma].
\end{equation}
Combing the tree-level Kähler potential with the superpotential and using the $\mathcal{N} = 1$ supergravity formula for the potential, we get that
\begin{equation}\label{8}
    V = e^K \Big( \sum_{a,b} g^{a \Bar{b}} D_a W \overline{D_b W} - 3|W|^2  \Big) \rightarrow e^K \Big( \sum_{i,j} g^{i \bar{j}} D_i W \overline{D}_j W \Big)
\end{equation}
where $D_a$ is the Kähler derivative $D_a W = \partial_a W  + W \partial_a K$ and $g_{a\bar{b}} = \partial_a \partial_{\bar{b}} K$ is the Kähler metric. Here $a$, $b$ are summed over all superfields and $i$, $j$ run over all moduli fields except $\rho$ as it cancels with the $-3|W|^2$ term. 

In order to find the minimum of the AdS potential we use the SUSY condition $D_{\rho}W =0$ which gives us
\begin{equation}\label{36}
   W_0 = -A e^{a \sigma} ( \frac{2}{3}a \sigma + 1).
\end{equation}
Substituting this into the minimum of the potential (\ref{8}) we get
\begin{equation}\label{37}
    V_{AdS} = -\frac{a^2 A^2 e^{-2a\sigma}}{6\sigma}
\end{equation}
where the volume modulus now has been stabilized while preserving SUSY.

\subsection{Uplifting AdS vacua}\label{upliftingadsvacua}

\begin{figure}[H]
\centering
\includegraphics[angle=0, width=0.9\linewidth]{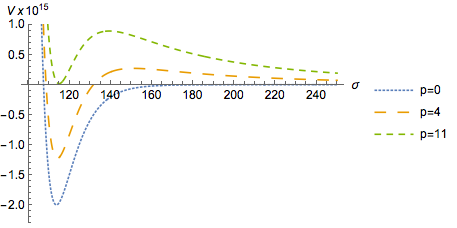}
\caption{Plot of the potential (multiplied by $10^{15}$), where $W_0 = -10^{-4}$, $A = 1$, $a = 0.1$ and the number of $\overline{D3}$-branes are  $p=0$ (blue),  $p=4$ (yellow) and $p=11$ (green).}
\label{fig.5}
\end{figure}

Now consider the case where too much flux is turned on so that the tadpole condition is satisfied when inserting one $\overline{D3}$-brane transverse to the compact manifold $M$.  The extra piece of energy from each $\overline{D}_3$-brane comes from the uplift term in the action (\ref{32}). KKLT reparametrize $e^{-12 u(x)}$ such that the volume modulus scales as $\frac{1}{\sigma^3}$: the volume of the compactification without warping is 
\begin{equation}\label{cyvolume}
    \int_M d^6 x h^{-1/4} g \sim e^{6u(x)} \alpha'^3.
\end{equation}
The added contribution thus yields
\begin{equation}\label{38}
    \delta V  = \frac{2 p}{g_s h_0 \sigma^3}
\end{equation} 
The potential now reads:
\begin{equation}\label{42}
    V^{KKLT}  = \frac{a A e^{-a\sigma} }{2\sigma^2} \Big( \frac{1}{3}\sigma aA e^{-a \sigma} + W_0 + A e^{-a \sigma} \Big)+ \frac{2 p}{g_s h_0 \sigma^3}  
\end{equation}
which have a non-negative minimum for $p \geq \frac{g_sh_0}{12}(A a \sigma e^{-a \sigma})^2$. The minimum is simply given by adding the energy contribution from the anti-branes to the AdS minimum (\ref{37}):
\begin{equation}\label{43}
    V_{min}^{KKLT} = -\frac{a^2 A^2 e^{-2a\sigma}}{6\sigma} + \frac{2 p}{g_s h_0 \sigma^3}
\end{equation}
In fig.\ref{fig.5}, the potential is plotted for a numerical example with three different numbers of included $\overline{D3}$-branes, with a minimum at $\sigma = 113.6$. 

Since we are dealing with type IIB compactified on a CY manifold, the string scale and 4-d Planck scale is related via the volume of the compact CY manifold, $V_{CY}$, given by (\ref{cyvolume}): 
\begin{equation}\label{plnckstringkklt}
    (m_{pl}^{(4)})^2 = \frac{2 V_{CY}}{g_s^2 (2 \pi)^7} \alpha^{\prime 4}= \frac{2\sigma^{3/2}}{g_s^2 (2 \pi)^7}m_s^2.
\end{equation}

The fact that the $\overline{D3}$
branes migrate to the tip of the throat assures that $\delta V$ will be exponentially suppressed. The warp factor at the tip of the throat is given by \cite{dorronsoro2017explicit}
\begin{equation}\label{warp}
    h_0 = \Big [ \Big( \frac{3}{2}\Big)^{1/3} b_0^2 g_s M \alpha' \epsilon ^{-4/3}  \Big] ^2
\end{equation}\label{40}
where the deformation parameter in the limit where $MK>>p$ is given by \cite{dorronsoro2017explicit}
\begin{equation}
    \epsilon =z = \Big( \frac{27}{4} \pi g_s \alpha'^2 KM \Big)^{3/8} \exp\Big(\frac{- \pi KM}{g_s M^2}\Big).
\end{equation}

Now we have presented the background of the minimized scalar potential in the KKLT construction that will be used to compute the AdS-lengths that goes into the expression for the cc in section.\ref{cosmologyonexpandingbubble}. However, we will also compute the minimized AdS scalar potential in the Large Volume Scenario, the most prominent alternative to KKLT, described in the next section.

 \subsection{The Large Volume Scenario}
 As was pointed out in the last section, there is much debate about the validity of the KKLT model.  The Large Volume Scenario (LVS) allows us to restore some of the issues with the model. The three main advantages LVS have over KKLT are: (a) there is no SUSY hierarchy problem, fluxes prefer $W_0 \sim 1$, (b) the SUSY breaking is not entirely due to the uplifting-mechanism so the effects from fluxes are not erased and (c) the volume is always exponentially large where the $\alpha'$ expansions are under better control due to the small expansion parameter, namely the inverse volume.  Another advantage of LVS is the lack of tachyonic direction in the scalar minimum, which often appear in KKLT solutions \cite{Landete_2016}.  Moreover, it seems like the LVS vacuum is self-consistent and safe from further corrections to the effective action. 

Now we will consider the simplest example of a LVS with with $\alpha'$ corrections to the Kähler potential (\ref{kähler}) and two Kähler moduli $\tau_b$ and $\tau$. We will also assume that one modulus takes an exponentially large value while the other modulus stays small. This setup will form the so called 'Swiss-cheese' structure of the CY three-form. 
In this picture, the volume modulus is given by \cite{dorronsoro2017explicit}
\begin{equation}\label{52}
    \mathcal{V} = \tau_{big}^{3/2} - \tau^{3/2} \approx  \tau_{big}^{3/2}.
\end{equation}

The volume modulus $\mathcal{V}$ is dimensionless and related to physical volumes via $\textit{vol}_6 \sim \mathcal{V}l_s^6 = \mathcal{V}(2 \pi)^6 \alpha'^3$. The string frame metric in this picture is given by

\begin{align}\label{53}
ds^2 &= \mathcal{V}^{1/3}e^{2 A(y)} ds_4^2 + e^{-2 A(y)}ds_{CY_0},\\
ds_{CY_0} &= \mathcal{V}^{-1/3} ds_{CY}^2 = g_{CY_0, mn}dy^m dy^n.
\end{align}
Here $ds_4^2$ is the 4-d Minkowski line element and the stabilized volume modulus can be identified with a constant shift of the warp factor \cite{Giddings_2006}:
\begin{equation}\label{55}
    e^{-4A(y)}= \mathcal{V}^{2/3} + h(y)
\end{equation}
where $h(y)$ describes the warping. 
The above ansatz can be used to find the relation between the string scale and Planck scale by reducing it to a 4-d Einstein frame \cite{dorronsoro2017explicit}:
\begin{equation}\label{56}
    \frac{m_{pl}^{(4)}}{2} = \frac{1}{2 (8 \pi G g_s)^2} \Big( \int d^6 y \sqrt{g_{CY_0}} e^{-4 A(y)} \mathcal{V}^{1/3} \Big) \approx \frac{(2\pi)^3 \alpha'^3 \mathcal{V}}{2 (8 \pi G g_s)^2} = \frac{\mathcal{V}}{2 \pi \alpha' g_s^2}.
\end{equation}

\begin{figure}[H]
  \center
  \subfloat[]{\includegraphics[width=0.43\textwidth]{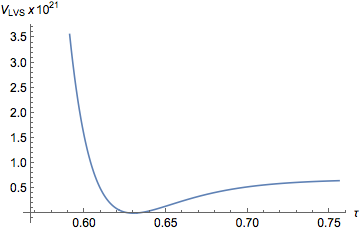}}
  \qquad \qquad
  \subfloat[]{\includegraphics[width=0.43\textwidth]{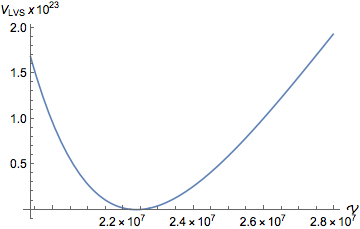}}
\caption{Potential against $\tau$ (a) and $\mathcal{V}$ (b) where the other direction were fixed using (\ref{59}) and (\ref{60}), respectively. The plots are made using the parameters given in Table \ref{table1}.}
\label{fig.8}
\end{figure}

Contributions to the superpotential  and Kähler potential takes the form
\begin{equation}\label{57}
   K =  -2 ln [ \mathcal{V} + \frac{\xi}{2} ] + ln [ \frac{g_s}{2} ]+ ln[-i \int \Omega \wedge \bar{\Omega}], \quad W = W_0 + \Sigma A_i e^{-a_i \tau_i /g_s}
\end{equation}
where $\frac{\xi}{2}$ comes from the leading $\alpha'$ correction and depends on the Euler number of the CY three-fold. Here $A_i$ are constants depending on the complex structure, and $a_i = 2\pi/N_i$ where $N_i = 1$ if the non-perturbative effect on a four-cycle $\tau_i$
arise from a Euclidean D3 brane, and $N_i = ND_7$ if the non-perturbative effect comes from gaugino condensation on a stack of D7 branes wrapped on $\tau_i$ \cite{dorronsoro2017explicit}. In the large volume limit where $\mathcal{V}^{3/2} \approx \tau_{big}^{3/2}$, the scalar potential is given by \cite{dorronsoro2017explicit}
\begin{equation}\label{58}
    V_{LVS} = \frac{g_s^4 (m_{pl}^{(4)})^4}{8 \pi} \Bigg[ \frac{2p}{g_s h_0 \mathcal{V}^{4/3}} + \frac{8g_s(aA)^2 \sqrt{\tau} e^{-2a \tau /g_s}}{3 g_s^2 \mathcal{V}} -  4aAW_0 \frac{\tau e^{-a \tau /g_s}}{g_s \mathcal{V}^2} + \frac{3 \xi W_0^2}{4 \mathcal{V}^3} \Bigg].
\end{equation}
To stabilize the moduli $\tau$ and $\mathcal{V}$ we find a stable minimum for the term in the square bracket. The factor outside the square bracket is necessary in order for the supergravity potential (\ref{8}) arising from the string frame LVS Kähler and superpotential (\ref{34}) to be consistent with dimensional reduction \cite{Anguelova_2009}. Due to non-perturbative effects and $\alpha'$ corrections, the scalar potential for $p=0$ have a non-SUSY AdS minimum where the volume is exponentially large. To find the minimum, we compute the first derivatives of the term in the square bracket of (\ref{58}):
\begin{align}\label{59} 
\partial_\tau V_{LVS} &= 0 \iff \mathcal{V}_{min} = \frac{3 e^{a\tau/g_s}g_s \sqrt{\tau}(a\tau -g_s)W_0}{aA (4a\tau -g_s)} \approx \frac{3 e^{a\tau/g_s}g_s \sqrt{\tau}W_0}{4aA} \\
\partial_{\mathcal{V}} V_{LVS} &= 0 \iff \tau_{min} = \frac{1}{9} \Big( \frac{27 \xi}{2} + \frac{16 p \mathcal{V}^{5/3}}{g_s h_0 W_0^2} \Big)^{2/3} \label{60}
\end{align}
where the approximate result in (\ref{59}) were used in (\ref{60}) for $\frac{a \tau}{g_s}>>1$.  This approximation should however be used with caution; substituting in the approximate result of (\ref{59}) and (\ref{60}) back into the LVS scalar potential (\ref{58}) for $p=0$ yields a a Minkowski vacuum (i.e. $V_{LVS}(\mathcal{V}_{min}, \tau_{min}) = 0$).  However, perturbing around (\ref{59}) and (\ref{60}) helps us find the true global minimum which we search for around the minima in fig.\ref{fig.8}.

Using the exact result in (\ref{59}) the minimized AdS potential takes the form
\begin{multline} \label{61}
    V_{LVS}^{min} = \frac{a^2 A^2 e^{-3 a \tau} g_s (m_{pl}^{(4)})^4}{288 h_0 \pi \tau^{3/2} (g_s -a \tau)^3W_0^2} \Bigg(-a A h_0 (g_s-4a \tau)W_0  \Big(16 \tau^{3/2}(a\tau -g_s)(2 a \tau + g_s) - \xi(g_s -4 a \tau)^2 \Big) \\ +    \frac{  3^{2/3}  8 a A p(g_s -4a \tau)^3 }{g_s W_0} \Big( \frac{e^{a \tau /g_s}g_s \sqrt{\tau}W_0 (a \tau-g_s}{{aA (4a\tau -g_s)}} \Big)^{5/3} \Bigg).
\end{multline}
Now, letting $\tau \rightarrow \tau_*$ denote the true minimized Kähler modulus, which we obtain numerically, and  substituting in the warp factor at the tip of the conifold (\ref{warp}) we get: 
\begin{multline}\label{62}
    V_{LVS}^{min} = \frac{a^3 A^3  \exp(-\frac{3a \tau_* }{g_s})(g_s - 4a \tau_*)}{288 g_s^2 \tau_*^{3/2} (g_s - a\tau_*) W_0^3} \Bigg( \frac{g_s^3 W_0^2}{\pi}\Big(-17 \tau_*^{3/2}(a\tau_* -g_s)(g_s + 2a \tau_*) + (g_s -4a \tau_*)^2 \xi \Big) \\ + 72 \times 2^{2/3}\exp\Big(\frac{-8 K \pi + 5a M \tau_*}{3 g_s M}\Big) \frac{g_s k p (g_s - 4 a \tau_*)^{1/3}}{b_0^4 M} \Big(\frac{g_s \sqrt{\tau_* (g_s-a\tau_*)W_0}}{aA} \Big)^{5/3} \Bigg).
\end{multline}

 \newpage
 
 \section{An effective cosmology on an expanding bubble}\label{cosmologyonexpandingbubble}
 
The KPV, KKLT and LVS constructions presented in the previous sections allow us to compute 4-d vacua. The KPV instanton described the decay of one such non-SUSY vacuum to another lower vacuum.  This decay takes place via the nucleation of a bubble that is made up of a spherical NS 5 domain wall which our universe is localized on. In this section, we will put the pieces together and compute properties of this braneworld and in particular the 3-d cc's.

 Cosmological properties on the bubble wall can be calculated by solving the Israel junction conditions (\ref{ijc}) across the boundary, which corresponds to Einstein’s field equations on the bubble. 
 The bulk metric $g_{ab}$ induce a metric on the bubble given by $h_{ab} = g_{ab} - N_a N_b$ where $N_a = N_a(x)$ is the unit vector normal defined in the propagating direction transverse to the bubble wall. The bulk metric is different inside and outside the bubble and the induced metric should be the same whether calculated with the bulk metric for either region.  The extrinsic curvature, $K_{\mu \nu}$, is defined as 
\begin{equation}\label{63}
    K_{\mu \nu} = N_{a;b} e^a_\mu e^b_\nu = h_a^{c} h_b^d \nabla_c N_d e^a_\mu e^b_\nu.
\end{equation}
where $e^a_\mu = \frac{\partial y^a}{\partial x^\mu}$ are tangent vectors with Latin indices labeling bulk-coordinates and Greek indices labelling coordinates on the bubble. By denoting the stress tensor on the bubble $t_{ab}$ and the trace of the extrinsic curvature as $K$, the Israel Junction condition can be written as
 \begin{equation}\label{ijc}
    - 8 \pi G t_{ab} = \Delta K_{ab} - \Delta K h_{ab}
\end{equation}
where we have summed over both sides of the bubble and let $\Delta K_{ab} = K^+_{ab} - K^-_{ab}$. To see an explicit dimensional dependence, this expression could be split into a trace and a trace-fee part so that 
\begin{equation}\label{75}
  8 \pi G\Big(  t_{ab}- \frac{1}{n-1}t h_{ab} \Big) =   \Delta K_{ab} - \frac{1}{n-1} \Delta K h_{ab}.
\end{equation}
Since the energy can flow from the hyper surface to the bulk, the energy momentum tensor is not necessarily conserved on the brane. The junction conditions will be used in the next subsection to find the tension and Friedmann equations of the expanding bubble.

 \subsection{AdS$_4$ decay to AdS$_4$}\label{adsdecaytoads}
Let's consider a $(2+1)$-dimensional bubble created through the nucleation of a non-SUSY $(3+1)$-dimensional AdS vacuum, decaying into a lower AdS vacuum. The weak gravity conjecture \cite{ooguri2017nonsupersymmetric} suggest that non-SUSY AdS vacua supported by fluxes must decay. However, even if this is true and the vacuum interior to the bubble would nucleate a second time, it will not affect this construction since it would never catch up and interact with the bubble wall as the surface is spacelike and asymptotes to the lightcone \cite{Banerjee_20191}. If an external bubble, a different universe, were to collide with the bubble from the second nucleation it could however potentially catch up to the bubble we live on.

\begin{figure}[H]
\centering
\includegraphics[angle=0, width=0.6\linewidth]{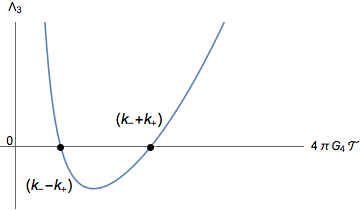}
\caption{The dimensionless 3d-cc vs the tension.}
\label{fig.9}
\end{figure}

Due to the spherical symmetry of the nucleated vacuum, the bulk metric in global coordinates can be written as 
\begin{equation}\label{bulkmetric}
    ds_\pm = -f_\pm(r)dt^2 + \frac{1}{f_\pm(r)}dr^2 + r^2 d\Omega_2^2
\end{equation}
with $\Omega_2 = d\theta^2 + \sin^2(\theta) d\phi^2$ being the usual two-sphere and $f_\pm(r) = 1+r^2 k_\pm ^2$ tells us if we are describing the interior (-) or the exterior (+) region of the bubble.  $k_\pm = \frac{1}{L_\pm}$ where $L_\pm$ is the curvature length of the spacetime in the respective regions. Since we are interested in dynamical solutions on the bubble, we define the coordinates there as $(R(\tau), T(\tau), \theta, \phi)$, where $\tau$ is the proper time as seen by an observer on the bubble. In terms of these coordinates, the normal vector in the propagating direction transverse to the bubble wall is given by
\begin{equation}\label{78}
    N_a = (-\dot{R}, \dot{T}, \theta, \phi). 
\end{equation}
where $\cdot = \frac{\partial}{\partial \tau}$.  By requiring that a point on the brane follows a timelike trajectory, the condition $g^{ab}N_a N_b = 1$ gives us the additional relation
\begin{equation}\label{79}
    \dot{T} = \frac{\sqrt{f(r)+\dot{R}^2}}{f(r)}.
\end{equation}
Now, the induced metric on the brane can be obtained by dividing and multiplying the right-hand side of the bulk-metric (\ref{bulkmetric}) with $d \tau^2$:
\begin{equation}\label{80}
\begin{split}
ds_{induced}^2 & = -f(r)dt^2 + \frac{1}{f(r)}dr^2 + r^2 d\Omega_2^2 \\
&= -f(r) \Big( \dot{T}^2 - \frac{\dot{R}^2}{f(r)^2} \Big)  d\tau^2 + R^2 d\Omega_2^2 \\ & =  -d\tau^2 + R^2 d\Omega_2^2
\end{split}
\end{equation}
where either choice of $\pm$ gives the same induced metric. Using (\ref{63}) we can compute the extrinsic curvature in the induced coordinates as
\begin{equation}\label{81}
    K_{\mu \nu} dx^\mu dx^\nu  =  -\frac{1}{f(R(\tau))\dot{T}}[ \Ddot{R} + \frac{1}{2}\frac{\partial f(R(\tau))}{\partial R} ] d\tau^2 +  R \dot{T} f(R(\tau)) d\Omega_2^2.
\end{equation}
The difference in extrinsic curvature across the bubble is given by 
\begin{equation}\label{82}
\begin{split}
\Delta K_{\mu \nu} dx^\mu dx^\nu = \Big( \frac{R k_-^2 + \Ddot{R}}{\sqrt{k_-^2 R^2 + \dot{R}^2}} - \frac{R k_+^2 + \Ddot{R}}{\sqrt{k_+^2 R^2 + \dot{R}^2}} \Big) d\tau^2 \\ + R^2 \Big( \sqrt{\frac{1 + \dot{R}^2}{R^2} + k_+^2} - \sqrt{\frac{1 + \dot{R}^2}{R^2} + k_-^2} \Big) d\Omega_2^2.
    \end{split}
\end{equation}
Now, solving the Israel junction conditions (\ref{ijc}) for the tension we get
\begin{equation}\label{83}
    \mathcal{T} = \frac{1}{4 \pi G_4} \Big( \sqrt{\frac{1 + \dot{R}^2}{R^2} + k_-^2} - \sqrt{\frac{1 + \dot{R}^2}{R^2} + k_+^2} \Big).
\end{equation}
Here we have assumed the matter on the bubble to be distributed as an isotropic perfect fluid so that the energy momentum tensor can be written as 
\begin{equation}
   t_{ab}= \text{diag}(-\mathcal{T}, \mathcal{T}, \mathcal{T}, \mathcal{T}).
\end{equation}
Solving the junction conditions for $\dot{R}$, we find the  Friedmann equation
\begin{equation}\label{84}
    H_3^2 = \Big(\frac{\dot{R}}{R}\Big)^2= -\frac{1}{R^2} + \frac{k_-^4 +(k_+^2 - 16 \pi^2 G_4^2 \mathcal{T}^2)^2 - 2k_-^2 (16 \pi^2 G_4^2 \mathcal{T}^2 + k_+^2)}{64 \pi^2 G_4^2 \mathcal{T}^2}.
\end{equation}
We will use the convention where $\Lambda_3$ is the constant part of the Friedmann equation and with mass dimension two:
\begin{equation}\label{85}
    \Tilde{\Lambda}_d = \Lambda_d (m_{pl}^{(d)})^{d-2}.
\end{equation} To be able to express $\Lambda_3$ in terms of the 3d-Planck mass, we will carry out a similar analysis as in \cite{Banerjee_20191} to find a relationship between $G_3$ and $G_4$ which in turn relates to the Planck mass via
\begin{equation}\label{86}
    G_d = \frac{1}{(m_{pl}^{(d)})^{d-2}}.
\end{equation}
Here $(d)$ in the right-hand side is not an exponent but denote the dimension of the Planck mass.
In fig.\ref{fig.9} the induced cc is plotted as a function of the tension and it becomes apparent that a small positive cc is expected near the critical tension.

By looking at the limit where the 3-d energy scales are much smaller than the 4-d energy scales: $\frac{1}{R}, \frac{\dot{R}}{R} << k_\pm$ the critical tension becomes
\begin{equation}\label{87}
    \mathcal{T}_{crit} = \frac{1}{4 \pi G_4}(k_- - k_+)
\end{equation}
which correspond to a Minkowski space. From this it becomes clear that the critical tension corresponds to the left root in fig.\ref{fig.9} and for the bubble to expand, we must have $\mathcal{T}< \mathcal{T}_{crit}$. So, letting $\mathcal{T} = \mathcal{T}_{crit}(1-\epsilon)$, and expanding in $\epsilon>0$ gives
\begin{equation}\label{88}
    H_3^2 = \Big( \frac{\dot{R}}{R}  \Big)^2 = -\frac{1}{R(\tau)^2} + 2 \epsilon k_-k_+ + \mathcal{O}(\epsilon^2).
\end{equation}
This can be rewritten as 
\begin{equation}\label{89}
    H_3^2 = -\frac{1}{R^2} + 4 \pi G_3 \Tilde{\Lambda}_3 + \mathcal{O}(\epsilon^2)
\end{equation}
where $G_3$ and $\Tilde{\Lambda}_4$ are identified as
\begin{equation}\label{90}
    G_3 = 2 G_4\frac{k_- k_+}{(k_- - k_+)}, \quad \Tilde{\Lambda}_3 = \mathcal{T}_{crit} - \mathcal{T}.
\end{equation} Using (\ref{86}) we thus get the relation between the 4-d and 3-d Planck mass as
\begin{equation}\label{91}
    \frac{1}{m_{pl}^{(3)}} = \frac{2}{(m_{pl}^{(4)})^2}\frac{k_- k_+}{(k_- - k_+)}
\end{equation}
The relation between $G_3$ and $G_4$ can be derived and verified by considering how a string pull and deform the brane, or gravitons propagating over the brane as explicitly done in \cite{Banerjee_2020} for $G_4$ and $G_5$ with the same result up to a dimensional-dependent numerical factor.

\subsubsection{AdS from LVS}\label{adsfromlvs}
When computing the cc it is important to distinguish between 4-d cc which is obtained from a minimized AdS vacuum and the 3-d cc which is fundamentally different and obtained from the Friedman equations. 

From (\ref{84}) we have that the 3-d cc, $\Lambda_3$, is given by:
\begin{equation}\label{92}
    \Lambda_3 =  \frac{k_-^4 +(k_+^2 - 16 \pi^2 G_4^2 \mathcal{T}^2)^2 - 2k_-^2 (16 \pi^2 G_4^2 \mathcal{T}^2 + k_+^2)}{64 \pi^2 G_4^2 \mathcal{T}^2}.
\end{equation}
The curvature lengths relate to the minimized scalar potential via 
\begin{equation}\label{93}
    \Lambda_d= V_{min} = \frac{(d-1)(d-2)}{2L^2} = \frac{(d-1)(d-2)}{2}k^2.
\end{equation}
For the LVS scenario, we thus get 
\begin{equation}\label{94}
k = \sqrt{-\frac{\Lambda_{4}}{3}} \frac{1}{m_{pl}} =  \sqrt{ - \frac{V_{min}^{LVS}}{3}} \frac{1}{m_{pl}} 
\end{equation}
Now, substituting $k_\pm$ into $\Lambda_3$ and using (\ref{91}) to get 3-d Planck mass units yields
\newpage

\begin{multline}
    \Lambda_3 = \frac{e^{-2a \tau_*/g_s}g_s \kappa_-^2 \kappa_+^2}{72b_0^6 h_0^2 M^3 \pi^4 \mathcal{V}_{min}^3 (\kappa_- - \kappa_+)^2} \Bigg[ 32a^2 A^2b_0^6 g_s h_0^2 M^3 \pi^3 \sqrt{\tau_*} \mathcal{V}_{min}^2  - 48aA b_0^6    e^{a\tau_* /g_s} g_s^2 h_0^3 M^3 \pi^3 \tau_* \mathcal{V}_{min} W_0^2  \\ + e^{2a \tau_* /g_s} \Big(\mathcal{V}_{min}^{10/3} p_-^2 - 2 \mathcal{V}_{min}^{5/3} p_- (-6b_0^6 g_s^2 h_0 M^3 \pi^3 + \mathcal{V}_{min}^{5/3}p_+) + (6b_0^6 g_s^2 h_0 M^3 \pi^3 + \mathcal{V}_{min}^{5/3}p_+)^2  \\ + 9b_0^6 g_s^3 h_0^2 M^3 \pi^3 \xi W_0^2 \Bigg] (m_{pl}^{(3)})^2
\end{multline}

where $\kappa$ is the dimensionless quantity of $k$:
\begin{equation}\label{105}
\kappa_\pm m_{pl}^{(4)} =  k_\pm .
\end{equation}

For the bubble to be energetically viable,  the vacuum associated with $k_+$ must be larger than $k_-$.  Therefore, $k_+(k_-)$ is obtained from $V_{min}^{LVS}$ assigned a larger (smaller) number of $\overline{D3}$-branes. 

For the parameter set in Table \ref{table1}, $\tau_*$ is stabilized at 
\begin{equation}\label{99}
   \tau_* \approx 0.642658
\end{equation}
and we get
\begin{align}\label{106}
    \mathcal{T}_{NS5} &= 2.72924 \times 10^{-12} (m_{pl}^{(3)})^3 \\
    k_+ &= 1.33074\times 10^{-24} m_{pl}^{(3)}\\
    k_- &= 1.33076\times 10^{-24} m_{pl}^{(3)}\\
    \Lambda_3  &=  6.84413 \times 10^{-41}(m_{pl}^{(3)})^2.
\end{align}
As expected, the cc is indeed small and positive showing that there exists a parameter space admitting a dS cosmology given that the instanton solution is valid which we will comment on in subsection \ref{Hierarchiesandscales}.

\begin {table}[H]
\caption {} \label{table1} 
\begin{center}
\begin{tabular}{ |c|c|c|c|c|c|c|c|c|c| } 
\hline
$p_-$  &  $p_+$ & M & a & $g_s$ & $\xi$ & A & K & $W_0$  & $\tau_{*}$  \\ 
\hline
1&  8 & 100 & $\pi$ & 0.1 & $1$ & 1.1 & 61 & 3 &0.642658  \\

\hline
\end{tabular}
\end{center}
\end{table}

\begin{figure}[H]

\subfloat[]{\includegraphics[width=0.48\textwidth]{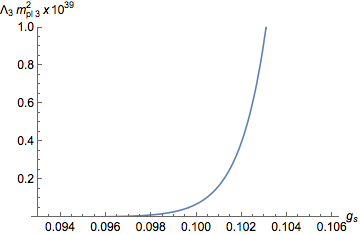}\label{10first}}
\hfill
\subfloat[]{\includegraphics[width=0.48\textwidth]{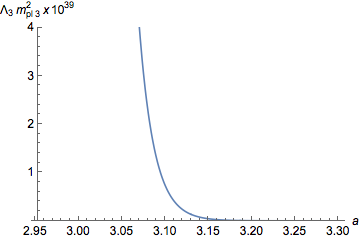}\label{10second}}

\subfloat[]{\includegraphics[width=0.48\textwidth]{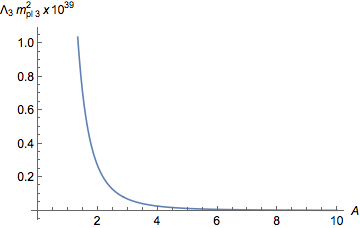}\label{10third}}
\hfill  
\subfloat[]{\includegraphics[width=0.48\textwidth]{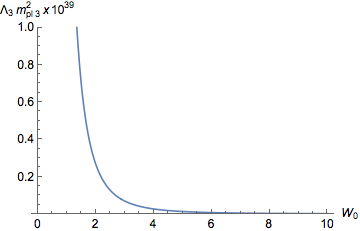}\label{10fourth}}

\subfloat[]{\includegraphics[width=0.48\textwidth]{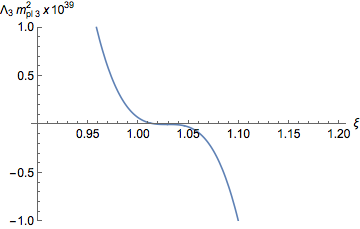}\label{10fifth}}
\hfill  
\subfloat[]{\includegraphics[width=0.48\textwidth]{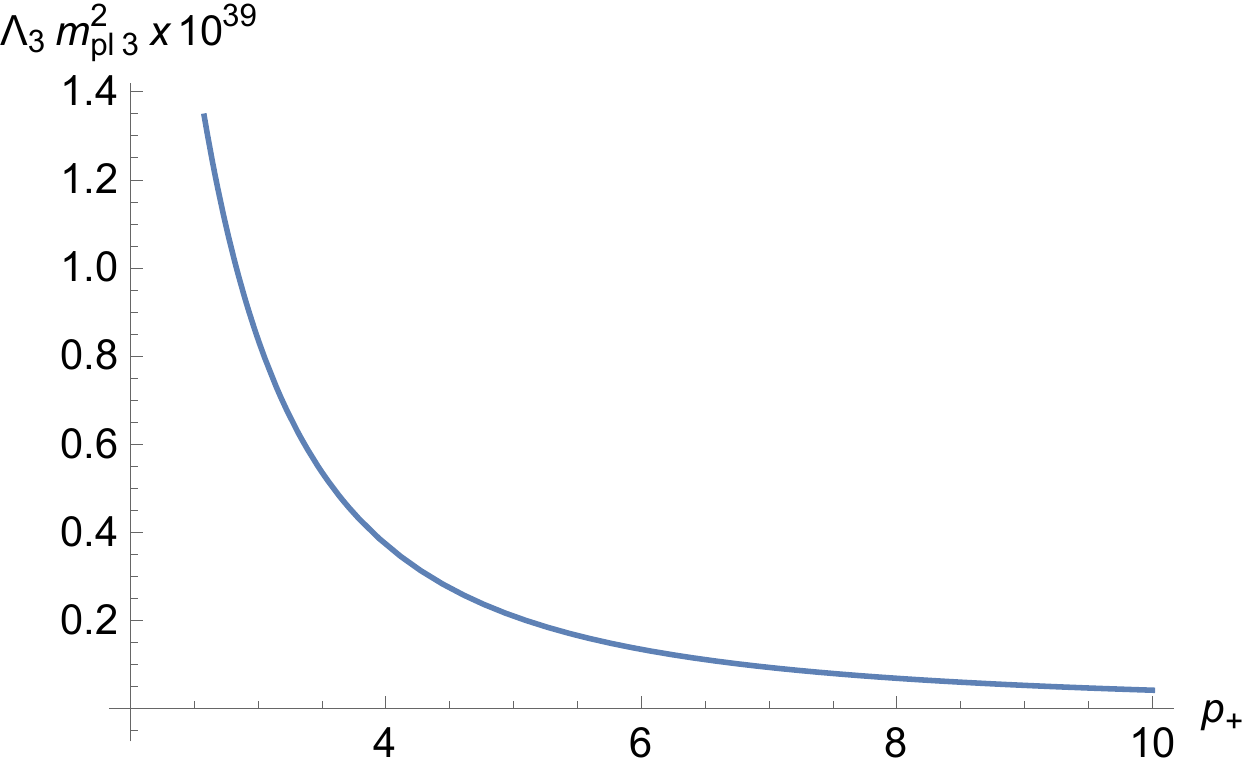} \label{10sixth}}

\caption{A plot of how the cosmological constant changes while varying the LVS parameters.}

\end{figure}

\newpage

We see from fig.\ref{10first} that we can move to a smaller coupling regime. However, moving to a larger regime will result in an exponential increase in the cc.  From fig.\ref{10second} and fig.\ref{10third} we see  that $a$ and $A$ should not be too many orders of magnitude away from $\pi$ and 1, respectively. In contrast to in the KKLT construction, we see in fig.\ref{10fourth} that the SUSY-breaking parameter $W_0$ cannot take values smaller than 1. We see from fig.\ref{10fifth} that the parameter $\xi$ should be close to 1 to avoid the cc to blow up. Finally in fig.\ref{10sixth} the cc is not sensitive to change is number of anti-branes for the decaying vacuum as long as the probe approximation $\frac{p}{M} \leq 8\%$ is satisfied.

\subsubsection{AdS from KKLT}\label{adsfromkklt}
Now let us consider the KKLT construction where the AdS lengths are given by
\begin{equation}\label{101}
k_\pm =  \sqrt{\frac{ - V_{KKLT}^{min}}{3}} \frac{1}{m^{(4)}_{pl}} = \frac{1}{3} \sqrt{  \Bigg( \frac{ a^2 A^2 e^{-2a \mathcal{T}}}{2 \mathcal{T}} - \frac{6 p_{\pm}  }{g_s h_0 \mathcal{T}^3} \Bigg)} \quad m^{(4)}_{pl}.
\end{equation}
Substituting this along with the NS5 domain wall tension (\ref{28}) into the expression for the cosmological constant we get
\begin{multline}\label{102}
\Lambda_3 = \frac{1}{72 \sigma^{9/2}}\Bigg[ -4a^2 A^2 e^{-2a \sigma} \sigma^{7/2} + \\  \frac{16 \sigma^3 p_-^2 + (3b_0^6 g_s^6 h_0M^3 \pi^2 + 4 \sigma^{3/2} p_+)^2 + 8 p_- (3b_0^6 g_s^6 h_0 M^3 \pi^2 \sigma^{3/2} -4\sigma^3 p_+)}{b_0^6 g_s^7 h_0^2 M^3\pi^2}  \Bigg] (m_{pl}^{(4)})^2
\end{multline}
where we have used (\ref{plnckstringkklt}) to relate the Planck mass and string mass.
Using the relation (\ref{86}) between the 4-d and 3-d Planck mass we get that 
\begin{multline}\label{104}
 \Lambda_3 = \frac{e^{- 6 a \sigma}}{5832 b_0^6 g_s^9 h_0^4 (\kappa_--\kappa_+)^2 M^3 \pi^2 \sigma^{21/2}} \Bigg[(-a^2 A^2 g_s h_0\sigma^2 +12 e^{2a \sigma}p_-)(-a^2A^2 g_s h_0 \sigma^2 +12e^{2a\sigma}p_+)  \\ (-4a^2 A^2 b_0^6 g_s^7 h_0^2 M^3 \pi^2 \sigma^{7/2}) + e^{2a\sigma}(16\sigma^3 p_-^2 + (3 b_0^6 g_s^6 h_0 M^3 \pi^2 + 4 \sigma^{3/2} p_+)^2 + \\ 8p_-(3b_0^6 g_s^6 h_0 M^3 \pi^2 \sigma^{3/2} -4\sigma^3 p_+) )
     \Bigg] (m_{pl}^{(3)})^2.
\end{multline}
Now, using the parameters in Table \ref{table2} yields the numerical values:  
\begin{align}\label{106}
    \mathcal{T}_{NS5} &= 3.73999 \times 10^{-26} (m_{pl}^{(3)})^3 \\
    k_+ &= 3.71881\times 10^{-15} m_{pl}^{(3)}\\
    k_- &= 4.75894\times 10^{-15} m_{pl}^{(3)}\\
    \Lambda_3  &= 8.83798 \times 10^{-23}(m_{pl}^{(3)})^2.
\end{align}

\begin {table}[H]
\caption {} \label{table2} 
\begin{center}
\begin{tabular}{ |c|c|c|c|c|c|c|c|c|c| } 
\hline
   $p_-$ & $p_+$ & M & a & $g_s$ & $\sigma$ & A & K & $W_0$ & $h_0$   \\ 
\hline
 1 & 4 & 200 & 0.1 & 0.05 & 113.58 & 1 & 36 & $-10^{-4}$ & $1.40435 \times 10^{11}$  \\
\hline

\end{tabular}
\end{center}
\end{table}

Since we have a completely analytical expression for the cc as seen by an observer on the bubble, we can again vary the parameters to get more insight into the parameter regime admitting a dS cosmology.

\begin{figure}[H]

\subfloat[]{\includegraphics[width=0.53\textwidth]{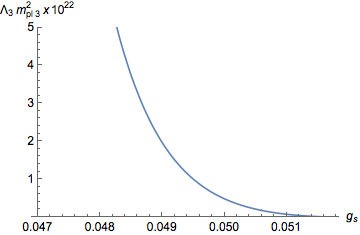}\label{first}}
\hfill
\subfloat[]{\includegraphics[width=0.53\textwidth]{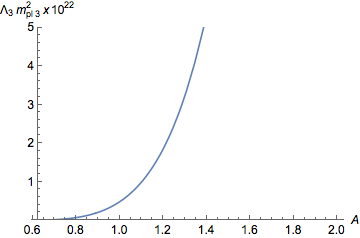}\label{second}}

\subfloat[]{\includegraphics[width=0.53\textwidth]{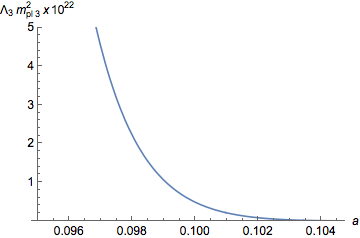}\label{third}}
\hfill  
\subfloat[]{\includegraphics[width=0.53\textwidth]{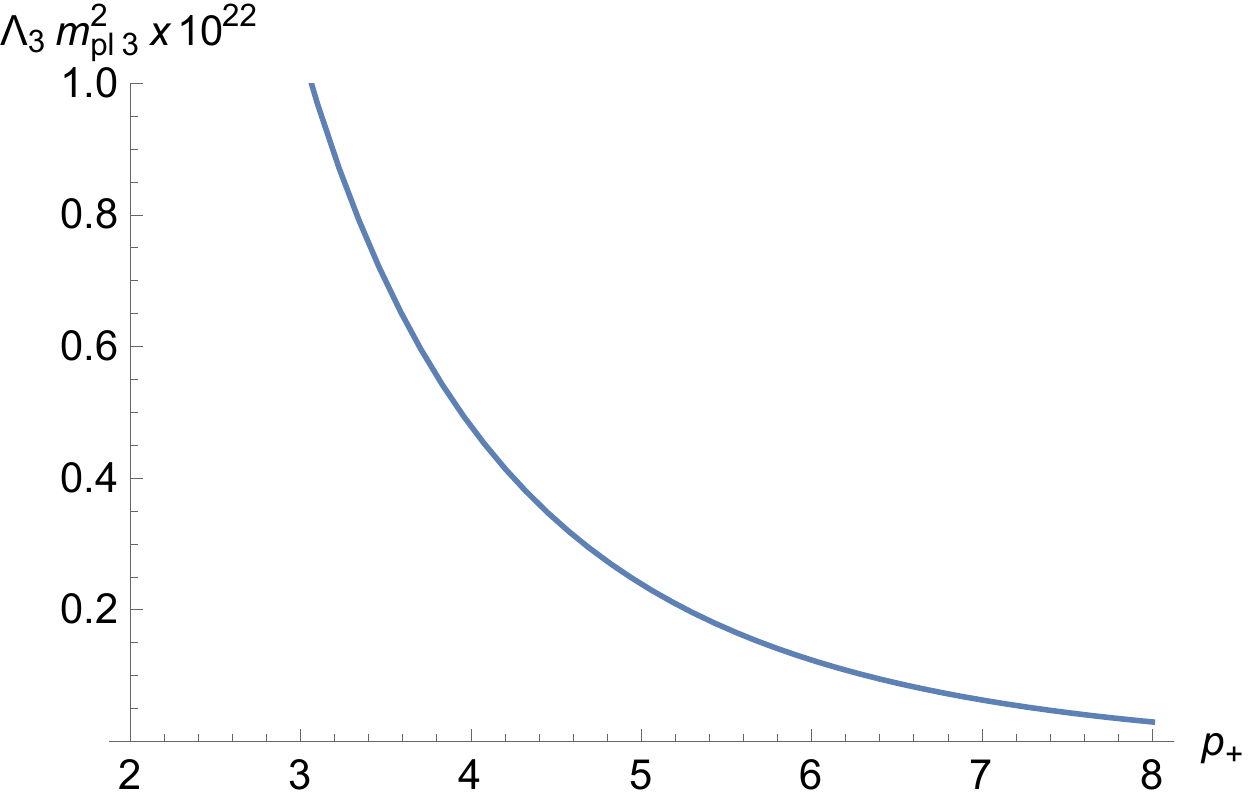}\label{fourth}}

\caption{A plot of how the cosmological constant changes while varying parameters in the KKLT construction.}

\end{figure}

\newpage
We have used that $g_s \sim \frac{1}{20}$ which is a relatively strong coupling. However, as illustrated in fig.\ref{first} a smaller coupling regime will lead to an exponential increase in the cc in contrast to the LVS case. In the same way as in the LVS, $a$ and $A$ are sensitive to change and a controlled calculation should have $a$ and $A$ not too many orders of magnitude away from  $\frac{1}{10}$ and $1$, respectively. As seen in fig.\ref{fourth}, we are free to pick any (although not arbitrary small) $p_+ <10$ without a drastic change in the magnitude of the cc.

\begin{figure}[H]
\centering
\includegraphics[angle=0, width=0.53\linewidth]{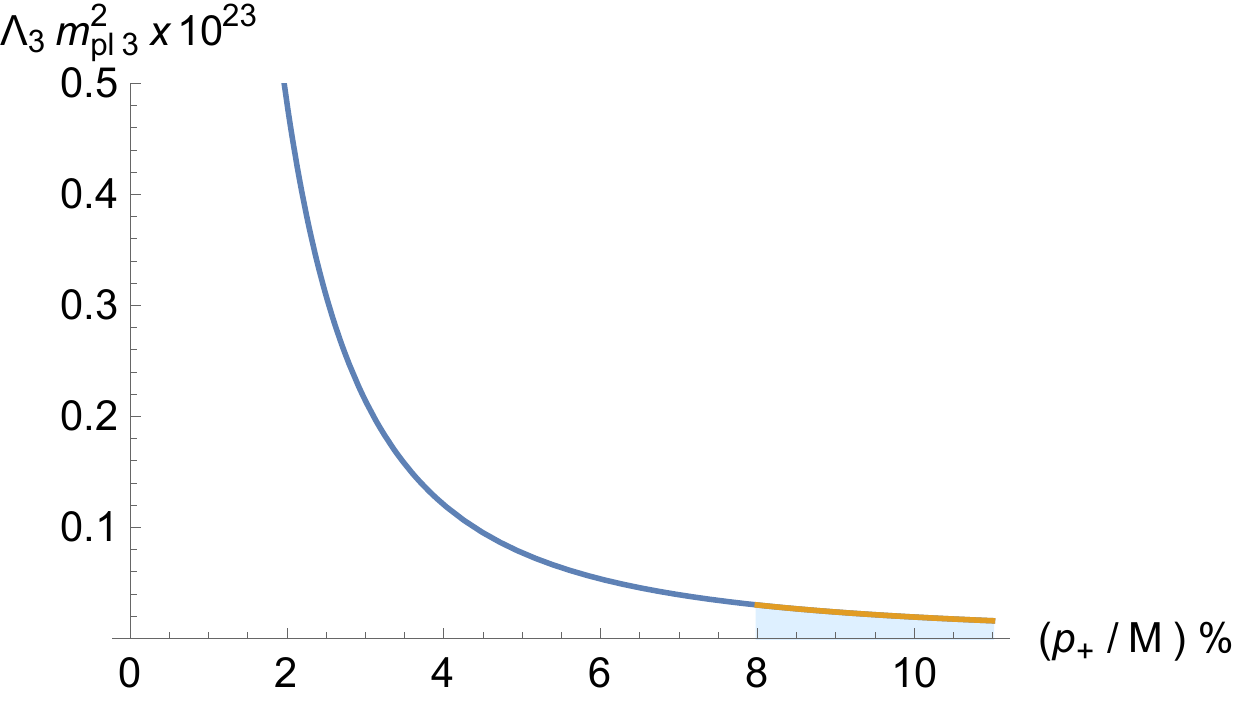}
\caption{The cosmological constant plotted against $\frac{p}{M}$. The shaded region is unstable with $\frac{p}{M}> 8 \%$}
\label{fig.11}
\end{figure}

In fig.\ref{fig.11} the cc is plotted against the ratio $\frac{p}{M}$ showing a runaway behaviour as  $\frac{p}{M}$ becomes arbitrary small.

\subsection{dS$_4$ decay to Minkowski$_4$}\label{dsdecaytominkowski}
Now consider the case where we have a dS spacetime decaying to a Minkowski spacetime, also via the nucleation of a bubble. As previously mentioned this model is not as theoretically motivated or as well understood as the AdS to AdS case. As in the previous subsection, we start from the metric \ref{bulkmetric}. Since there is no curvature length in Minkowski space we get $f_+ = 1$ and by analytically continue the AdS length to a dS length: $k \rightarrow ik$, we get $f_- = 1-k^2r^2$. In this metric we have that $r \rightarrow \rho$. Repeating the same process as in the previous sub section yields a tension
\begin{equation}\label{117}
  \mathcal{T}_f = \frac{\sqrt{1+ \rho'^2} - \sqrt{1-k^2\rho^2 + \rho'^2}}{4 \pi G \rho}.
\end{equation}
The Friedmann equation in terms of the tension then becomes 
\begin{equation}\label{118}
   H_3^2 =  -\frac{1}{\rho^2} + \frac{(k^2 + 16 \pi^2 G^2 \mathcal{T}_f^2)^2}{64 \pi^2 G_4^2 \mathcal{T}_f^2}.
\end{equation} 

Our bubble is realized by an instanton solution where the sphere nucleates
at rest at a finite size due to energy conservation. The radius of the instanton is such that the energy liberated when reducing the cc balances the energy cost of the tension of the bubble. At the instant when the bubble is formed, where $\dot{R}=0$, we have the radius
\begin{equation}\label{119}
    R = \rho = \sqrt{\frac{1}{\Lambda_3}}.
\end{equation}
We also have that $H_3^2 \approx \Lambda_3$ in the late time limit (i.e. for large $\rho$ so that the curvature term vanish). Using these ingredients while requiring the tension to be real gives the condition \label{120}
\begin{equation}
    H_3 > H_4 \Longleftrightarrow \rho < L_4.
\end{equation}
Since the nucleation event is non-local and the bubble forms as a spacelike surface at a single instant, one might think that it doesn't matter if it extends beyond the dS horizon, but the above condition tells us that the bubble must fit inside the dS horizon.

\begin{figure}[H] \label{fig.12}
  \subfloat[]{\includegraphics[width=0.53\textwidth]{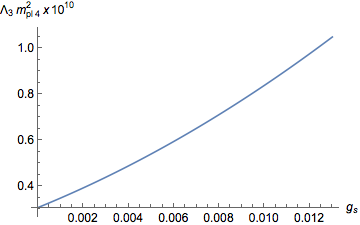}}
  \hfill     
  \subfloat[]{\includegraphics[width=0.53\textwidth]{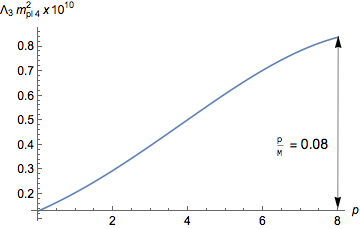}}\label{fig.12 (a)}
 
  \center
  \subfloat[The region to the left of the arrow shows the regime where $\frac{p}{M}> 8 \%$]{\includegraphics[width=0.53\textwidth]{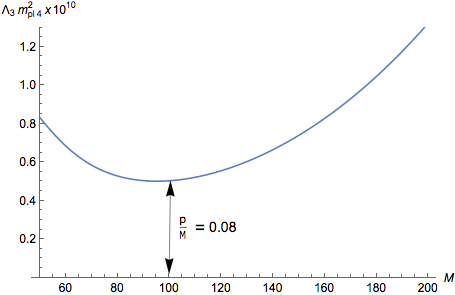}}\label{fig.12 (b)}
\caption{Plots showing how sensitive the cosmological constant is to change in certain parameters. }
\label{fig12}
\end{figure}

Now, from (\ref{118}) we obtain the cosmological constant 
\begin{equation}\label{121}
    \Lambda_3 = \frac{(k^2 + 16 \pi^2 G^2 \mathcal{T}_f^2)^2}{64 \pi^2 G_4^2 \mathcal{T}_f^2}.
\end{equation}
Using (\ref{101}) and the minimized potential (\ref{27}) from KPV, the four-dimensional Hubble constant in the late-time limit is given by 
\begin{equation}\label{122}
   H_{4,KPV} =   \sqrt{\frac{V_{KPV}^{min}}{3}}\frac{1}{m_{pl}^{(3)}} \approx \sqrt{\frac{\frac{\mu_3 p}{g_s} \Big(1-\frac{8\pi^2 p^2}{3b_0^{12}M^2} \Big)}{3}}\frac{1}{m_{pl}^{(3)}}.
\end{equation}
Now, letting $H_4 \rightarrow H_{4,KPV}$ and substituting in the NS 5-brane tension (\ref{28}) into $\mathcal{T}_f$ the cosmological constant yields
\begin{equation}\label{123}
 \Lambda_3 = \frac{( 9b_0^{18} g_s M^5 + 6 \times 10^3 b_0^{12} M^2  p \pi - 16 \times 10^3 p^3 \pi^3)^2}{5184 \times 10^9 b_0^{30} M^7 \pi^4}.
\end{equation}
Letting the string coupling be $g_s = 0.01$, $M=100$ and $p=8$ such that we are in the limit $\frac{p}{M} \approx 8 \%$ gives the numerical value
\begin{equation}\label{124}
    \Lambda_3 \approx 6.916 \times 10^{-11} (m_{pl}^{(4)})^2.
\end{equation}
In fig.\ref{fig12} the cc is plotted against parameters to see how sensitive they are to change.
As seen in the above figures the cc does not change by more than one order of magnitude in the plotted ranges. So, the parameters could vary by several orders of magnitude, while $\frac{p}{M} \leq 8\%$ is satisfied, without drastically change the magnitude of the cc. It is however important to note that the cc is in terms of the 4-d Planck mass and thus not according to an observer localized at the bubble. Expressing the cc in terms of the 3-d Planck mass requires the relationship between $G_4$ and $G_3$ which is an open problem as it is unknown whether gravity is localized on the brane in this picture. Finding this relation would lead to a nontrivial rescaling of the $\Lambda_3$-axis which might change the behaviour of the graphs.

\subsection{Hierarchies and scales}\label{Hierarchiesandscales}
In both the LVS and KKLT case in the AdS to AdS decay we have
\begin{equation}\label{equality1}
m_{pl}^{(3)} > m_{pl}^{(4)} > k > \Lambda_3
\end{equation}
as required. If $m_{pl}^{(3)}$ is not greater than $m_{pl}^{(4)}$, the corrections to the effective 3-d gravity will be large and the interpretation of the junction equation as a 3-d Friedmann equation may not be valid.   If $m_{pl}^{(4)}$ is not greater than $k$ then the 4-d AdS is not well described by general relativity and the use of the Israel junction conditions are not valid. For the $\alpha'$ corrections to the compact 6-d manifold to be under control we also need  $m_{pl}^{(3)} >m_{pl}^{(4)} > m_s$. It follows from these inequalities that 
\begin{equation}\label{inequality2}
  (m_{pl}^{(4)})^3  > (m_{pl}^{(3)})^3 > \mathcal{T}_{NS5}.
\end{equation}
However, this hierarchy should not be too big since then the energy scale of the tension would have an energy scale of the same order as low-energy 3-d observers which would rule out anything like "normal" braneworlds because observers, e.g. people living on the brane, could alter it just by walking around. The cosmology from the KKLT potentials satisfy (\ref{inequality2}) with a reasonable hierarchy. However, this is not the case for the LVS where more fine-tuning of the parameters to generate a smaller tension might be required. Due to the open problem of localizing gravity in the case where dS decays to Minkowski it is too early to conduct the above analysis in this picture. 

In the AdS to AdS channel, we must also have 
\begin{equation}\label{ineq}
\frac{\mathcal{T}_{crit}}{\mathcal{T}_{NS5}} = 1 + \epsilon
\end{equation}
for the expansion (\ref{89}) to be valid. A numerical check with the parameters in Table \ref{table1} and Table \ref{table2} shows that $ \mathcal{T}_{NS5}>>\mathcal{T}_{crit}$ for both the KKLT and LVS case, thus more fine-tuning is necessary for an energetically viable model with a valid instanton solution.

\newpage
\section{Outlook and discussion}\label{outlookanddiscussion}
In conclusion we explicitly computed a 3-d effective dS cosmology in the braneworld construction by \cite{Banerjee_2018}. 
This approach is fundamentally different from finding dS as minimized scalar potential via a reliable uplifting mechanism of an AdS vacuum. Rather, we derive the Friedmann equations and ask what cc an observer living on the bubble wall would see. It is however still necessary to minimize the AdS potentials to be substituted into the cc obtained from the Friedmann equations. To stabilize all moduli we must work in a regime where the parameters and supergravity approximations are valid and thus fine-tuning is still required.  Hence, the notion of multiverses and anthropic selection has to be taken seriously. 

The cc has been computed in (2+1)-d and will therefore not correspond to the cc measured in the universe we live in with (3+1) noncompact dimensions. However, since the dS Swampland conjectures state that no dS should be found in any dimension, this example is equally important. 

In the dS to Minkowski channel, we find a parameter set admitting a dS cosmology. However, for the AdS to AdS channel to be energetically viable, the ratio between critical tension (\ref{87}) and the SUSY NS 5-domain wall tension (\ref{28}) must be greater than and close to one. But, we find that $\mathcal{T}_{NS5}>>\mathcal{T}_{crit}$ so the bubble does not nucleate with the parameters in this paper and further fine-tuning is hence necessary. Even though the particular parameter sets explored for the AdS to AdS channel does not allow the bubble to nucleate, the braneworld model in this paper remain a promising new way of looking at the universe as an effective dS cosmology is realized automatically and there is thus no need for a fundamental dS vacuum. It also offers a natural explanation of dark energy using higher dimensional physics. 

Active work is being done to repeat this process of finding an effective dS in one higher dimension. If an effective dS cosmology were to be computed in 4-d, the tension should be well above any energy density probed by the LHC or high energy astrophysical processes. The challenge is finding a decaying 5-d AdS that is perturbatively stable without any tachyonic directions. In recent work a systematic search of the supergravity landscape was done using machine learning \cite{Bobev_2020}. Several new critical points of a superpotential corresponding to non-SUSY AdS vacua were found, however, all with Tachyonic direction. Once a 5-d AdS that only decays due to none-perturbative effects is found and an explicit effective 4-d dS cosmology can be generated, the roam of testing the braneworld construction and string theory using cosmological observations would open up.

\acknowledgments

 I would like to express my sincere gratitude to Marjorie Schillo for supervising my master thesis from which this article emerged from. I would also like to offer my special thanks to Ulf Danielsson and Suvendu Giri for insightful comments and discussions.

\newpage
\bibliographystyle{JHEP}
\bibliography{bib}

\end{document}